\font\tenfrakturb=eufb10
\font\tenfraktur=eufm10
\font\tenmsbm=msbm10
\font\sevenfrakturb=eufb7
\font\sevenfraktur=eufm7
\font\sevenmsbm=msbm7
\font\fivefrakturb=eufb5
\font\fivefraktur=eufm5
\font\fivemsbm=msbm5
\newfam\bgothicfam
\newfam\gothicfam
\newfam\msbmfam
\textfont\bgothicfam = \tenfrakturb \scriptfont\bgothicfam=\sevenfrakturb
\scriptscriptfont\bgothicfam=\fivefrakturb
\textfont\gothicfam = \tenfraktur \scriptfont\gothicfam=\sevenfraktur
\scriptscriptfont\gothicfam=\fivefraktur
\textfont\msbmfam = \tenmsbm \scriptfont\msbmfam=\sevenmsbm
\scriptscriptfont\msbmfam=\fivemsbm

\def\Bbb{\tenmsbm\fam\msbmfam}

\catcode`@=11
\def\renewcounter#1{\@definecounter{#1}\@ifnextchar[{\@newctr{#1}}{}}
\documentstyle[twoside]{article}
\catcode`\@=11
\long\def\@makefntext#1{
\protect\noindent \hbox to 3.2pt {\hskip-.9pt  
$^{{\eightrm\@thefnmark}}$\hfil}#1\hfill}		

\def\@makefnmark{\hbox to 0pt{$^{\@thefnmark}$\hss}}	
	
\def\ps@myheadings{\let\@mkboth\@gobbletwo
\def\@oddhead{\hbox{}
\rightmark\hfil\eightrm\thepage}   
\def\@oddfoot{}\def\@evenhead{\eightrm\thepage\hfil
\leftmark\hbox{}}\def\@evenfoot{}
\def\sectionmark##1{}\def\subsectionmark##1{}}
\oddsidemargin=\evensidemargin
\newcounter{sectionc}\newcounter{subsectionc}\newcounter{subsubsectionc}
\renewcommand{\section}[1] {\vspace{12pt}\addtocounter{sectionc}{1} 
\setcounter{subsectionc}{0}\setcounter{subsubsectionc}{0}\noindent 
{\tenbf\thesectionc. #1}\par\vspace{5pt}}
\renewcommand{\subsection}[1] {\vspace{12pt}\addtocounter{subsectionc}{1} 
\setcounter{subsubsectionc}{0}\noindent 
{\bf\thesectionc.\thesubsectionc. {\kern1pt \bfit #1}}\par\vspace{5pt}}
\renewcommand{\subsubsection}[1]{\vspace{12pt}\addtocounter{subsubsectionc}{1}
\noindent{\tenrm\thesectionc.\thesubsectionc.\thesubsubsectionc.
{\kern1pt \tenit #1}}\par\vspace{5pt}}
\newcommand{\nonumsection}[1] {\vspace{12pt}\noindent{\tenbf #1}
\par\vspace{5pt}}
\newcounter{appendixc}
\newcounter{subappendixc}[appendixc]
\newcounter{subsubappendixc}[subappendixc]
\renewcommand{\thesubappendixc}{\Alph{appendixc}.\arabic{subappendixc}}
\renewcommand{\thesubsubappendixc}
{\Alph{appendixc}.\arabic{subappendixc}.\arabic{subsubappendixc}}
\renewcommand{\appendix}[1] {\vspace{12pt}
        \refstepcounter{appendixc}
        \setcounter{figure}{0}
        \setcounter{table}{0}
        \setcounter{lemma}{0}
        \setcounter{theorem}{0}
        \setcounter{corollary}{0}
        \setcounter{definition}{0}
        \setcounter{equation}{0}
        \renewcommand{\thefigure}{\Alph{appendixc}.\arabic{figure}}
        \renewcommand{\thetable}{\Alph{appendixc}.\arabic{table}}
        \renewcommand{\theappendixc}{\Alph{appendixc}}
        \renewcommand{\thelemma}{\Alph{appendixc}.\arabic{lemma}}
        \renewcommand{\thetheorem}{\Alph{appendixc}.\arabic{theorem}}
        \renewcommand{\thedefinition}{\Alph{appendixc}.\arabic{definition}}
        \renewcommand{\thecorollary}{\Alph{appendixc}.\arabic{corollary}}
        \renewcommand{\theequation}{\Alph{appendixc}.\arabic{equation}}
        \noindent{\tenbf Appendix \theappendixc #1}\par\vspace{5pt}}
\newcommand{\subappendix}[1] {\vspace{12pt}
        \refstepcounter{subappendixc}
        \noindent{\bf Appendix \thesubappendixc. {\kern1pt \bfit #1}}
\par\vspace{5pt}}
\newcommand{\subsubappendix}[1] {\vspace{12pt}
        \refstepcounter{subsubappendixc}
        \noindent{\rm Appendix \thesubsubappendixc. {\kern1pt \tenit #1}}
\par\vspace{5pt}}
\newcommand{\textlineskip}{\baselineskip=13pt}
\newcommand{\smalllineskip}{\baselineskip=10pt}
\def\eightcirc{
\begin{picture}(0,0)
\put(4.4,1.8){\circle{6.5}}
\end{picture}}
\def\eightcopyright{\eightcirc\kern2.7pt\hbox{\eightrm c}} 
\newcommand{\copyrightheading}[1]
{\vspace*{-2.5cm}\smalllineskip{\flushleft
{\footnotesize International Journal of Modern Physics A, #1}\\
{\footnotesize $\eightcopyright$\, World Scientific Publishing
 Company}\\
 }}

\newcommand{\pub}[1]{{\begin{center}\footnotesize\smalllineskip 
Received #1\\
\end{center}
}}

\def\abstracts#1#2#3{{
\centering{\begin{minipage}{4.5in}\baselineskip=10pt\footnotesize
\parindent=0pt #1\par 
\parindent=15pt #2\par
\parindent=15pt #3
\end{minipage}}\par}} 
\def\keywords#1{{
\centering{\begin{minipage}{4.5in}\baselineskip=10pt\footnotesize
{\footnotesize\it Keywords}\/: #1
 \end{minipage}}\par}}
\newcommand{\bibit}{\nineit}
\newcommand{\bibbf}{\ninebf}
\renewenvironment{thebibliography}[1]
{\frenchspacing
 \ninerm\baselineskip=11pt
 \begin{list}{\arabic{enumi}.}
{\usecounter{enumi}\setlength{\parsep}{0pt}
 \setlength{\leftmargin 12.7pt}{\rightmargin 0pt} 
 \setlength{\itemsep}{0pt} \settowidth
{\labelwidth}{#1.}\sloppy}}{\end{list}}
\newcounter{itemlistc}
\newcounter{romanlistc}
\newcounter{alphlistc}
\newcounter{arabiclistc}

\newcommand{\fcaption}[1]{
        \refstepcounter{figure}
        \setbox\@tempboxa = \hbox{\footnotesize Fig.~\thefigure. #1}
        \ifdim \wd\@tempboxa > 5in
           {\begin{center}
        \parbox{5in}{\footnotesize\smalllineskip Fig.~\thefigure. #1}
            \end{center}}
        \else
             {\begin{center}
             {\footnotesize Fig.~\thefigure. #1}
              \end{center}}
        \fi}
\newcommand{\tcaption}[1]{
        \refstepcounter{table}
        \setbox\@tempboxa = \hbox{\footnotesize Table~\thetable. #1}
        \ifdim \wd\@tempboxa > 5in
           {\begin{center}
        \parbox{5in}{\footnotesize\smalllineskip Table~\thetable. #1}
             \end{center}}
        \else
             {\begin{center}
             {\footnotesize Table~\thetable. #1}
              \end{center}}
        \fi}
\def\@citex[#1]#2{\if@filesw\immediate\write\@auxout
{\string\citation{#2}}\fi
\def\@citea{}\@cite{\@for\@citeb:=#2\do
{\@citea\def\@citea{,}\@ifundefined
{b@\@citeb}{{\bf ?}\@warning
{Citation `\@citeb' on page \thepage \space undefined}}
{\csname b@\@citeb\endcsname}}}{#1}}
\newif\if@cghi
\def\cite{\@cghitrue\@ifnextchar [{\@tempswatrue
\@citex}{\@tempswafalse\@citex[]}}
\def\citelow{\@cghifalse\@ifnextchar [{\@tempswatrue
\@citex}{\@tempswafalse\@citex[]}}
\def\@cite#1#2{{$\null^{#1}$\if@tempswa\typeout
{IJCGA warning: optional citation argument 
ignored: `#2'} \fi}}

\def\pmb#1{\setbox0=\hbox{#1}
\kern-.025em\copy0\kern-\wd0
\kern.05em\copy0\kern-\wd0
\kern-.025em\raise.0433em\box0}

\def\fnt#1#2{\footnotetext{\kern-.3em
{$^{\mbox{\scriptsize #1}}$}{#2}}}
\def\fpage#1{\begingroup
\voffset=.3in
\thispagestyle{empty}\begin{table}[b]\centerline{\footnotesize #1}
\end{table}\endgroup}
\def\runninghead#1#2{\pagestyle{myheadings}
\markboth{{\protect\footnotesize\it{\quad #1}}\hfill}
{\hfill{\protect\footnotesize\it{#2\quad}}}}
\font\tenrm=cmr10
\font\tenit=cmti10 
\font\tenbf=cmbx10
\font\bfit=cmbxti10 at 10pt
\font\ninerm=cmr9
\font\nineit=cmti9
\font\ninebf=cmbx9
\font\eightrm=cmr8

\def\bh{${\Bbb R}^2\times {\Bbb S}^2\>$}

\def\qed{\hbox{${\vcenter{\vbox{  
   \hrule height 0.4pt\hbox{\vrule width 0.4pt height 6pt
   \kern5pt\vrule width 0.4pt}\hrule height 0.4pt}}}$}}

\begin{document}
\runninghead{A. A. Bytsenko \& Yu. P. Goncharov}
{ Dirac Monopoles and Hawking Radiation in Kottler Spacetime}
\normalsize\textlineskip
\thispagestyle{empty}
\setcounter{page}{1}
\copyrightheading{Vol. 17, No. 32 (2002) 4947--4957}
\vspace*{0.88truein}
\fpage{1}
\centerline{\bf DIRAC MONOPOLES AND HAWKING RADIATION}
\vspace*{0.035truein}
\centerline{\bf IN KOTTLER SPACETIME}
\vspace*{0.37truein}
\centerline{\footnotesize A. A. BYTSENKO\footnote{On leave of absence from 
Sankt-Petersburg State Polytechnical University, Russia}}
\vspace*{0.015truein}
\centerline{\footnotesize\it Departamento de Fisica, Universidade Estadual
de Londrina}
\baselineskip=10pt
\centerline{\footnotesize\it Caixa Postal 6001, Londrina-Parana, Brazil}
\vspace*{10pt}
\centerline{\footnotesize YU. P. GONCHAROV}
\vspace*{0.015truein}
\centerline{\footnotesize\it Theoretical Group,
Experimental Physics Department, State Polytechnical University}
\baselineskip=10pt
\centerline{\footnotesize\it Sankt-Petersburg 195251, Russia}
\vspace*{10pt}
\vspace*{0.225truein}
\pub{20 July 2002}
\vspace*{0.21truein}
\abstracts{
  The natural extension of Schwarzschild metric to the case of
nonzero cosmological constant $\Lambda$ known as the Kottler metric is
considered and it is discussed under what circumstances the given metric
could describe the Schwarzschild black hole
immersed in a medium with nonzero energy density. Under the latter situation
such an object might carry topologically inequivalent configurations of
various fields. The given possibility is analysed for complex scalar field
and it is shown that the mentioned configurations might be tied with natural
presence of Dirac monopoles on black hole under consideration. In turn, this
could markedly modify the Hawking radiation process.
}{}{}
\vspace*{10pt}
\vspace*{1pt}\textlineskip 
\section{Introductory remarks} 
\vspace*{-0.5pt}
\noindent
At present time there are certain indications that formation and evolution
of a significant part of black holes occur in active galactic nuclei
(see, e .g. Ref.\cite{EZ99} and references therein). Under the circumstances
the notion of black hole as an isolated object being described, for example,
by the Schwarzschild (SW) metric evidently becomes inadequate to the situation.
In active galactic nuclei the regions filled with a nonzero energy density
or strong magnetic fields are the rule rather than exception. As a result,
the properties of black holes may be changed under these conditions and to
take into account this change one should modify the description of black
holes. One can note that solutions of the Einstein equations which can model
the mentioned conditions are known for a long time. In particular, the
Kottler metric which can be written down in the form

$$ds^2=g_{\mu\nu}dx^\mu\otimes dx^\nu\equiv
adt^2-a^{-1}dr^2-r^2(d\vartheta^2+\sin^2\vartheta d\varphi^2) \eqno(1)$$
with $a=1-2M/r+\Lambda r^2/3$
is the solution of the Einstein equations with cosmological constant $\Lambda$
($M$ is the black hole mass) and is known since 1918
(see, e. g., Refs.\cite{Exsol}). Inasmuch as the
cosmological constant $\Lambda$ can be associated with the nonzero energy
density (in usual units) $\rho=\Lambda c^4/(8\pi G)$ (see e. g. Ref. 3)
then the
metric (1) may be interpreted as the one describing the SW black hole
immersed in a medium with the nonzero energy density. It should be noted,
however, that metric (1) is not asymptotically flat one and the given
circumstance can lead to difficulties of formulating the scattering problem
necessary for the Hawking radiation analysis (see below).

On the other hand, the nontrivial topological properties of spacetimes
may play essential role while studying quantum geometry of fields on them
(see, e. g., our reviews\cite{GB}).
The black hole spacetimes are not exceptions. Really,
as was shown in a recent series of the papers
\cite{{Gon94},{Gon96},{Gon97},{Gon99}}, the black holes can
carry the whole spectrum of topologically inequivalent configurations (TICs)
for miscellaneous fields, in the first turn, complex scalar and spinor ones.
The mentioned TICs can essentially modify the Hawking radiation from black
holes\cite{GF} and also can give some possible statistical explanation of
the black hole entropy\cite{GonN}. Physically, the existence of TICs should
be obliged to the natural presence of magnetic U(N)-monopoles (with $N\ge1$)
on black holes though the total (internal) magnetic charge (abelian or
nonabelian) of black hole remains equal to zero.

In present paper we would like to clear up to what extent the mentioned
features can be inherent to the black holes described by metric (1). For
to simplify calculations our considerations will restrict
themselves to the TICs of complex scalar field and, accordingly, to the
Dirac U(1)-monopoles. Sec. 2 serves for obtaining the auxiliary relations
necessary for the rest of paper. Sec. 3 is devoted to description of
Dirac monopoles on the black holes under exploration and also to separation
of variables for the corresponding wave equations while Sec. 4 contains
the correct statement of the scattering problem for scalar particles
interacting with the mentioned Dirac monopoles and description of the
conforming $S$-matrix necessary to analyse the Hawking radiation process.
The latter is discussed in Sec. 5 and Sec. 6 gives concluding remarks.

  Throughout the paper we employ the system of units with $\hbar=c=G=1$,
unless explicitly stated. Finally, we shall denote $L_2(F)$ the set of
the modulo square integrable complex functions on any manifold $F$
furnished with an integration measure.

\section{Preliminary considerations}
First of all it should be noted that the standard spacetime topology on which
the metric (1) with arbitrary $a=a(r)$ can be realized in a natural way is of
\bh-form. In the given paper we shall, however, deal only with the Kottler
metric where the form of $a$ was fixed above in Sec. 1.
Let us introduce a number of quantities that for metric (1) with
$\Lambda\ne0$ will
generalize the corresponding ones when $\Lambda=0$, i. e., in the SW black
hole case. Besides we shall consider $\Lambda>0$ within the paper.

Let us consider the equation $a(r)=1-2M/r+\Lambda r^2/3=0$ that can be
rewritten as $ r^3+3r/\Lambda-6M/\Lambda=0$. Discriminant of the latter
equation is equal to $D=(1/\Lambda)^3+(3M/\Lambda)^2 >0$ which means the
mentioned equation to have one real and two complex conjugated roots.
The only real root is
$$r_+=\frac{1}{\sqrt{\Lambda}}(\root3\of{\alpha}+\root3\of{\beta})\eqno(2)$$
with $\alpha=3M\sqrt{\Lambda}-\sqrt{1+9M^2\Lambda}$,
$\beta=3M\sqrt{\Lambda}+\sqrt{1+9M^2\Lambda}$ and we have the expansion
$$a(r)=\frac{\Lambda}{3r}(r-r_+)
\left(r^2+r_+r+\frac{6M}{\Lambda r_+}\right)\>.\eqno(3)$$
It is clear that $r_+\to2M$
when $\Lambda\to 0$ and we can see that horizon of the SW black hole is
shifted to $r_+$ at $\Lambda>0$. Under this situation when defining the
black hole entropy as $S=\pi r_+^2$ we shall find the black hole temperature
$T$ from the relation $1/T=\partial S/\partial M$ which entails
$$T=\frac{\Lambda\sqrt{1+9M^2\Lambda}\>(\alpha\beta)^{2/3}}
{2\pi(\alpha^{1/3}+\beta^{1/3})[3M\Lambda(\alpha^{2/3}-\beta^{2/3})+
\sqrt{\Lambda(1+9M\Lambda^2}(\alpha^{2/3}+\beta^{2/3})]}\>, \eqno(4)$$
so when $\Lambda\to0$, $T\to 1/8\pi M$, the SW black hole temperature.

Further, we define the variable $r_*$ by the relation
$$r_*=\int\frac{dr}{a}=\frac{r_+^2}{2(3M-r_+)}\ln{\left(\frac{r}{r_+}-1\right)}
-\frac{r_+^2}{4(3M-r_+)}\ln{\left[1+\frac{\Lambda r_+}{6M}(r^2+rr_+)\right]}$$
$$+\left[\frac{6M}{\Lambda(3M-r_+)}+\frac{r_+^3}{2(3M-r_+)}\right]
\frac{1}{\sqrt{\frac{24M}{\Lambda r_+}-r_+^2}}
\arctan\frac{2r+r_+}{\sqrt{\frac{24M}{\Lambda r_+}-r_+^2}}-\frac{r_+}{2}\>,
\eqno(5)$$
where we choose the integration constant for the expression (5) to pass on to
the standard Regge-Wheeler variable $r_*= r+2M\ln[r/(2M)-1]$ at $\Lambda\to0$.
At last, let us introduce the dimensionless quantities $x=r_*/M$, $y=r/M$,
$y_+=r_+/M$
so that for them we shall have
$$x=\frac{y_+^2}{2(3-y_+)}\ln{\left(\frac{y}{y_+}-1\right)}
-\frac{y_+^2}{4(3-y_+)}\ln{\left[1+\frac{\Lambda M^2y_+}{6}(y^2+yy_+)\right]}$$
$$+\left[\frac{6}{\Lambda M^2(3-y_+)}+\frac{y_+^3}{2(3-y_+)}\right]
\frac{1}{\sqrt{\frac{24}{\Lambda M^2y_+}-y_+^2}}
\arctan\frac{2y+y_+}{\sqrt{\frac{24}{\Lambda M^2y_+}-y_+^2}}-\frac{y_+}{2}\>,
\eqno(6)$$
i. e, $y=y(x)$ is a reverse function to (6) and again at $\Lambda\to0$
the relation is simplified to $x=y+2\ln(0.5y-1)$. At this point we face the
difficulty that at $y\to y_+$, $x\to-\infty$ while when $y\to+\infty$, $x\to$
some finite value which is different from the SW black hole case, where
$y_+=2\le y\le +\infty$, whereas $-\infty<x<+\infty$. Under this situation
the correct statement of the scattering problem necessary for the
Hawking radiation analysis is impossible so long as the latter problem should
be considered on the whole $x$-axis\cite{Fir99}.

To overcome this obstacle it should be noted that the region in universe
with the nonzero energy density (and where the SW black hole can be located)
has, generally speaking, some finite sizes. Under the circumstances, if $r_0$
is such a characteristic size then the metric (1) should describe the
black hole at $r_+\le r\le r_0$, while at $r_0\le r\le+\infty$ the black hole
should be described by metric (1) with $\Lambda=0$, i. e., by the standard
SW metric. As a result, the relation (6) should be applied
when $-\infty < x\le x_0$, otherwise, at $x_0\le x < +\infty$
the corresponding $y(x)$ should be determined from the equation
$x=y+2\ln(0.5y-1)$ and one should put $y_0=r_0/M$ while
$x_0=y_0+2\ln(0.5y_0-1)$.

 Now we shall have a function $y(x)$ defined on the
whole $x$-axis and the conforming scattering problem can be posed correctly
(see Sec. 4). It is clear the size $r_0$ should be specified under the
concrete conditions so in what follows we shall use it as a parameter
in further considerations.

\section{Dirac monopoles and separation of variables}

As was disscussed in Refs.\cite{{Gon94},{Gon96},{Gon97}},
the black holes can carry TICs of a complex scalar field that
are conditioned by the availability of a countable
number of complex line bundles over the \bh-topology underlying
the 4D black hole physics. Each TIC corresponds to sections of a complex
line bundle $E$ while the latter can be characterized by its Chern number
$n\in \Bbb{Z}$
(the set of integers). TIC with $n=0$ can be called {\it untwisted},
while the rest of the TICs with
$n\not=0$
should be referred to as {\it twisted}.
Using the fact that all the mentioned line bundles can be trivilized over
the chart of local coordinates
$(t,r,\vartheta,\varphi) $
covering almost the whole manifold \bh
one can obtain a suitable wave equation on the given chart for TIC $\phi$
with mass $\mu_0$ and Chern number $n\in\Bbb{Z}$
in the form
$$|g|^{-1/2}
(\partial_\mu-ieA_\mu)[g^{\mu\nu}(|g|)^{1/2}
(\partial_\nu-ieA_\nu)\phi]=-\mu_0^2\phi\>,\eqno(7)$$
where the specification
depends on the gauge choice for connection $A_\mu$
(vector-potential for the corresponding Dirac monopole) in the line bundle $E$
with the Chern number $n$. We shall here consider the gauge
where one can put the above connection 1-form equal to
$A= A_\mu dx^\mu=-\frac{n}{e}\cos\vartheta d\varphi$.
Then the curvature of the bundle $E$ is $F=dA=
\frac{n}{e}\sin\vartheta d\vartheta\wedge d\varphi$. We can further introduce
the Hodge star operator on 2-forms $F$ of any $k$-dimensional
(pseudo)riemannian manifold
$B$ provided with a (pseudo)riemannian metric $g_{\nu\tau}$ by the relation
(see, e. g., Ref.\cite{Bes87})
$$F\wedge\ast F=(g^{\nu\alpha}g^{\tau\beta}-g^{\nu\beta}g^{\tau\alpha})
F_{\nu\tau}F_{\alpha\beta}
\sqrt{|g|}\,dx^1\wedge dx^2\cdots\wedge dx^k \eqno(8)$$
in local coordinates $x^\nu$. In the case of the metric (1) this yields
$$\ast(dt\wedge dr)=\sqrt{|g|}g^{tt}g^{rr}d\vartheta\wedge d\varphi=
-r^2\sin\vartheta d\vartheta\wedge d\varphi\>,$$
$$\ast(d\vartheta\wedge d\varphi)=\sqrt{|g|}
g^{\vartheta\vartheta}g^{\varphi\varphi}dt\wedge dr=
\frac{1}{r^2\sin\vartheta}dt\wedge dr\>,\eqno(9)$$
so that $\ast^2=\ast\ast=-1$, as should be for the manifolds with lorentzian
signature\cite{Bes87}. After this we find
$\ast F= \frac{n}{er^2}dt\wedge dr\>,$
and integrating over the surface $t= const$, $r=const$ with topology
${\Bbb S}^2$ gives rise to the Gauss theorem
$$\int_{S^2}\ast F=0 \eqno(10)$$
and the Dirac charge quantization condition
$$\int_{S^2} F=4\pi\frac{n}{e}=4\pi q \eqno(11)$$ with magnetic charge $q$.
Besides, the Maxwell equations $dF=0$, $d\ast F =0$ are clearly fulfilled with
the exterior differential $d=\partial_t dt+\partial_r dr+
\partial_\vartheta d\vartheta+\partial_\varphi d\varphi$ in coordinates
$t,r,\vartheta,\varphi$.
Also it should be emphasized that the total
(internal) magnetic charge $Q_m$ of black hole which
should be considered as the one summed up over all the monopoles remains
equal to zero because

$$Q_m=\frac{1}{e}\sum\limits_{n\in{\Bbb{Z}}}\,n=0\>,\eqno(12)$$
so the external observer does not see any magnetic charge of the black
hole though the monopoles are present on black hole in the sense
described above.

\subsection{Monopole masses}

We can estimate the corresponding monopole masses so long as we are
effectively in the asymptotically flat spacetime due to the remarks made
late in Sec. 2 and one can use the $T_{00}$-component
of the energy-momentum tensor for electromagnetic field

$$T_{\mu\nu}={1\over4\pi}(-F_{\mu\alpha}F_{\nu\beta}g^{\alpha\beta}+
{1\over4}F_{\beta\gamma}F_{\alpha\delta}\,g^{\alpha\beta}g^{\gamma\delta}
g_{\mu\nu})\eqno(13)$$
to find the monopole masses according to

$$m_{\rm{mon}}(n)=\int\limits_{{\Bbb R}\times {\Bbb S}^2}\,
T_{00}\sqrt{\gamma}d^3x
=\int\limits_{{\Bbb R}\times {\Bbb S}^2}
T_{00}\,r^2\sin\vartheta\,a^{-1/2}d^3x \eqno(14)$$
with $a(r)$ of (1) and $\sqrt{\gamma}d^3x=r^2\sin\vartheta a^{-1/2}d^3x$ is
the volume element for the spatial part (with topology
${\Bbb R}\times{\Bbb S}^2$) of the metric (1)
while $T_{00}$-component is proved to be equal to
$\frac{1}{16\pi}
aF^2_{\vartheta\varphi}g^{\vartheta\vartheta}g^{\varphi\varphi}$ with
$F_{\vartheta\varphi}=n\sin\vartheta/e$. This entails

$$m_{\rm{mon}}(n)=\frac{n^2}{4e^2}\left[\int\limits_{r_+}^{r_0}
\sqrt{1-\frac{2M}{r}+\frac{\Lambda r^2}{3}}\frac{dr}{r^2}+
\int\limits_{r_0}^\infty\sqrt{1-\frac{2M}{r}}\frac{dr}{r^2}\right] $$
$$=\frac{n^2}{4e^2}\left[\int\limits_{r_+}^{r_0}
\sqrt{1-\frac{2M}{r}+\frac{\Lambda r^2}{3}}\frac{dr}{r^2}+
\frac{1}{3M}-\frac{1}{3M}\left(1-\frac{2M}{r_0}\right)^{3/2}\right]\>.
\eqno(15) $$
It is clear that at $r_0\to+\infty$, $m_{\rm mon}(n)\to\infty$ and
the conforming Compton wavelength
$\lambda_{\rm mon}=1/m_{\rm mon}\to0$, i. e., $\lambda_{\rm mon}<< r_+$
and the given monopoles might reside in black hole under discussion as
quantum objects.

\subsection{Separation of variables}

If returning to the Eq. (7), then, in the way analogous to that
of Refs.\cite{{Gon94},{Gon96},{Gon97}}, one can show the
equation (7) to have
in $L_2($\bh$)$
a complete set of solutions of the form
$$f_{\omega n m l}=\frac{(2\pi\omega)^{-1/2}}{r}e^{i\omega t}
Y_{nlm}(\vartheta,\varphi)R_{\omega nl}(r),\qquad\qquad
l=|n|,|n|+1,...,\qquad|m|\leq l,\eqno(16)$$
where the explicit form of the {\it monopole (spherical) harmonics}
$Y_{nlm}(\vartheta,\varphi)$ can be found in Refs.
\cite{{Gon94},{Gon96},{Gon97}}.
As to the function $R_{\omega nl}(r)$, then it obeys the equation
$$a\partial_rar^2\partial\frac{R_{\omega nl}}{r}+\omega^2rR_{\omega nl}=
a\left[\mu_0^2+\frac{l(l+1)-n^2}{r^2}\right]rR_{\omega nl}\>, \eqno(17)$$
which can be rewritten as
$$\frac{d^2R_{\omega nl}}{dr_*^2}+(\omega^2-\mu_0^2)R_{\omega nl}=
\left[\mu_0^2(a-1)+\frac{a}{r}\frac{da}{dr}+
a\frac{l(l+1)-n^2}{r^2}\right]R_{\omega nl}\eqno(18)$$
with the help of (5). At last, when denoting $k=M\omega$ and introducing
the functions $\psi
(x,k,n,l)=R_{\omega nl}[My(x)]$,
where $y(x)$ is a function reverse to (6),
we shall obtain that $\psi(x,k,n,l)$
obeys the Schr\"odinger-like equation
$$\left[\frac{d^2}{dx^2}+(k^2-\mu_0^2M^2)\right]\psi(x,k,n,l)=
V(x,M,\Lambda,n,l)\psi(x,k,n,l)\>, \eqno(19)$$
where
$$V(x,M,\Lambda,n,l)=
\left[1-\frac{2}{y(x)}+\frac{\Lambda M^2}{3}y^2(x)\right]
\left[\frac{2}{y^3(x)}+\frac{2\Lambda M^2}{3}
+\frac{l(l+1)-n^2}{y^2(x)}\right]$$
$$+(\mu_0M)^2\left[\frac{\Lambda M^2}{3}y^2(x)
-\frac{2}{y(x)}\right]\>, \> -\infty\le x \le x_0\>,
\eqno(20)$$
and
$$V(x,M,\Lambda,n,l)=
\left[1-\frac{2}{y(x)}\right]
\left[\frac{2}{y^3(x)}+\frac{l(l+1)-n^2}{y^2(x)}\right]
-(\mu_0M)^2\frac{2}{y(x)}\>, \> x_0\le x \le +\infty\>. \eqno(21)$$

Now we can discuss how to pose the correct scattering problem for
Eq. (19) because the corresponding solutions of this problem will be
just necessary to analyse the Hawking radiation from black hole.

\section{The scattering problem}

One can notice that potential $V(x,M,\Lambda,n,l)$ of (20)--(21) satisfies
the conditions explored in Refs.\cite{Fir99} and referring for more details
to those works we may here use their results to correctly formulate the
scattering problem for Eq. (19). Namely, the correct statement of the
scattering problem will consist in searching for two
solutions $\psi^+(x,k^+,n,l)$,
$\psi^-(x,k,n,l)$ of the equation (19)
obeying the following conditions

$$\psi^+(x,k^+,n,l)=
\cases{e^{ikx}+
s_{12}(k,l,n)e^{-ikx}+o(1),&$x\to-\infty$,\cr
s_{11}(k,l,n)w_{-i\mu,\frac{1}{2}}(-2ik^+x)+o(1),&$x\to+\infty$,\cr}$$
$$\psi^-(x,k,n,l)=\cases{s_{22}(k,l,n)e^{-ikx}
+o(1),&$x\to-\infty$,\cr
w_{i\mu,\frac{1}{2}}(2ik^+x)+
s_{21}(k,l,n)w_{-i\mu,\frac{1}{2}}(-2ik^+x)+o(1),&$x\to+\infty$,\cr}
\eqno(22)$$
where
$k^+(k)=\sqrt{k^2-(\mu_0M)^2}$,
$\mu=\frac{(\mu_0M)^2}{(k^+)^2}=\frac{(\mu_0M)^2}{k^2-(\mu_0M)^2}$
and the functions
$w_{\pm i\mu,\frac{1}{2}}(\pm z)$ are related to the Whittaker
functions $W_{\pm i\mu,\frac{1}{2}}(\pm z)$ (concerning the latter
ones see e. g. Ref.\cite{Abr64}) by the relation
$$w_{\pm i\mu,\frac{1}{2}}(\pm z)=
W_{\pm i\mu,\frac{1}{2}}(\pm z)e^{-\pi\mu/2}\>,$$
so that one can easily gain asymptotics (using the corresponding ones for
Whittaker functions\cite{Abr64})

$$w_{i\mu,\frac{1}{2}}(-2ik^+x)=
e^{ik^+x}e^{i\mu\ln|2k^+x|}
[1+O(|k^+x|^{-1})], \> x\to+\infty\>,$$
$$w_{-i\mu,\frac{1}{2}}(2ik^+x)=e^{-ik^+x}
e^{-i\mu\ln|2k^+x|}
[1+O(|k^+x|^{-1})], \> x\to+\infty\>.\eqno(23)$$

We can see that there arises some $S$-matrix with elements
$s_{ij}, i, j = 1, 2$
and it satisfies certain unitarity relations (for more details see
Refs.\cite{Fir99}) As will be seen below, for calculating the Hawking
radiation we need the coefficient $s_{11}$, consequently, we need to have
some algorithm for numerical computation of it inasmuch as the latter cannot
be evaluated in exact form. The given algorithm can be extracted from the
results of Refs.\cite{Fir99} To be more precise
$$s_{11}(k,l,n)=2ik/[f^-(x,k,l,n),f^+(x,k^+,l,n)]\>,\eqno(24)$$
where [,] signifies the Wronskian of functions $f^-,f^+$, the
so-called Jost type solutions of Eq. (19).
In their turn, these functions and
their derivatives obey the certain integral equations. Since the Wronskian
does not depend on $x$ one can take the following form of the mentioned
integral equations

$$f^-(x_0,k,l,n)=e^{-ikx_0}+\frac{1}{k}\int\limits^{x_0}_{-\infty}
\sin[k(x_0-t)]V^-(t,l,n)f^-(t,k,l,n)dt\>,\eqno(25)$$
$$(f^-)'_x(x_0,k,l,n)=-ike^{-ikx_0}+
\int\limits^{x_0}_{-\infty}\cos[k(x_0-t)]V^-(t,l,n)f^-(t,k,l,n)dt\>,
\eqno(26)$$

$$f^+(x_0,k^+,l,n)=w_{i\mu,\frac{1}{2}}(-2ik^+x_0)+$$
$$\frac{1}{k^+}
\int\limits_{x_0}^{+\infty}{\rm Im}[w_{i\mu,\frac{1}{2}}(-2ik^+x_0)
w_{-i\mu,\frac{1}{2}}(2ik^+t)]
V^+(t,l,n)f^+(t,k^+,l,n)dt\>,\eqno(27)$$
$$(f^+)'_x (x_0,k^+,l,n)=\frac{d}{dx}w_{i\mu,\frac{1}{2}}(-2ik^+x_0)+$$
$$\frac{1}{k^+}\int\limits^{+\infty}_{x_0}{\rm Im}[\frac{d}{dx}
w_{i\mu,\frac{1}{2}}(-2ik^+x_0)
w_{-i\mu,\frac{1}{2}}(2ik^+t)]V^+(t,l,n)f^+(t,k^+,l,n)dt\>,
\eqno(28)$$
where the potentials
$$V^-(x,l,n)=
\left[1-\frac{2}{y(x)}+\frac{\Lambda M^2}{3}y^2(x)\right]
\left[(\mu_0M)^2+\frac{2}{y^3(x)}+\frac{2\Lambda M^2}{3}
+\frac{l(l+1)-n^2}{y^2(x)}\right]\>,\eqno(29)$$
$$V^+(x,l,n)=2(\mu_0M)^2\left[\frac{1}{x}-\frac{1}{y(x)}\right]+
\left[1-\frac{2}{y(x)}\right]
\left[\frac{l(l+1)-n^2}{y^2(x)}+\frac{2}{y^3(x)}\right]\>\eqno(30)$$
tend to 0 when, respectively, $x\to-\infty$ or $x\to+\infty$.
The relations (24)--(30) can be employed for numerical calculation of
$s_{11}$ ( see Refs.\cite{GF} for the case $\Lambda=0$).

To summarize, we can now obtain
the general solution of (19) in the class of
functions restricted on the whole $x$-axis in the linear combination form

$$\psi(x,k,n,l)=C^1_{nl}(k)\psi^+(x,k^+,n,l)+
C^2_{nl}(k)\psi^-(x,k,n,l)\>,\eqno(31)$$
since a couple of the functions $\psi^+(x,k^+,n,l)$, $\psi^-(x,k,n,l)$
forms a fundamental system of solutions for the equation (19).

Having obtained all
the above, we can discuss the Hawking radiation process for any TIC of complex
scalar field.

\section{Modification of Hawking radiation}
  As can be seen, the equation (7) corresponds to the lagrangian
$${\cal L}=(-g)^{1/2}(g^{\mu\nu}\overline{{\cal D}_\mu\phi}
{\cal D}_\nu\phi-{\mu_0}^2\overline{\phi}\phi)\>,\eqno(32)$$
where the overbar signifies complex conjugation, ${\cal D}_\mu=\partial_\mu-
ieA_\mu$ and, as a result, we have the conforming energy-momentum
tensor for TIC with the Chern number $n$

$$T_{\mu\nu}={\rm Re}[(\overline{{\cal D}_\mu\phi})({\cal D}_\nu\phi)
-\frac{1}{2}g_{\mu\nu}g^{\alpha\beta}(\overline{{\cal D}_\alpha\phi})
({\cal D}_\beta\phi)-\mu_0^2g_{\mu\nu}\overline{\phi}\phi]\>.\eqno(33)$$

When quantizing twisted TIC with the Chern number $n$ we shall
take the set of functions
$f_{\omega nlm}$ of (16) as a basis in $L_2$(\bh) so that
$|m|\leq l$, $l=|n|,|n|+1$,... and we shall
normalize the monopole spherical harmonics
$Y_{nlm}(\vartheta,\varphi)$ to 1, i. e., by the condition
$$\int\limits_0^\pi\,\int\limits_0^{2\pi}
\overline{Y_{nlm}}(\vartheta,\varphi)
Y_{nl^\prime m^\prime}(\vartheta,\varphi)
\sin\vartheta d\vartheta d\varphi=\delta_{ll^\prime}
\delta_{mm^\prime}\>.\eqno(34)$$

After this we can evidently realize the procedure of quantizing
TIC with the Chern number $n$, as usual, by expanding this TIC in the modes
$f_{\omega nlm}$

$$\phi=\sum\limits_{l=|n|}^\infty\sum\limits_{|m|\leq l}
\int\limits_{\mu_0}^\infty\,d\omega
(a^-_{\omega nlm}f_{\omega nlm}+
b^+_{\omega nlm}\overline{f_{\omega nlm}})\,,$$
$$\phi^+=\sum\limits_{l=|n|}^\infty\sum\limits_{|m|\leq l}
\int\limits_{\mu_0}^\infty\,d\omega
(b^-_{\omega nlm}f_{\omega nlm}+
a^+_{\omega nlm}\overline{f_{\omega nlm}})\,,\eqno(35)$$

so that $a^\pm_{\omega nlm}$, $b^\pm_{\omega nlm}$ should be interpreted
as the corresponding
creation and annihilation operators for charged scalar particle in both the
gravitational field of black hole and the field of the conforming monopole
with the Chern number $n$ and we have the standard commutation relations
$$[a^-_i,\,a^+_j]=\delta_{ij},\>[b^-_i,\,b^+_j]=\delta_{ij}\>
\eqno(36)$$
and zero for all other commutators,
where $i =\{\omega nlm\}$ is a generalized index.
  Under the circumstances one can speak about the Hawking radiation
process for any TIC of complex scalar
field. Actually, one should use the energy-momentum tensor of (33)
for TIC with the Chern number $n$ in quantum form
$$T_{\mu\nu}={\rm Re}[(\overline{{\cal D}_\mu}\phi^+)({\cal D}_\nu\phi)
-{1\over2}g_{\mu\nu}g^{\alpha\beta}(\overline{{\cal D}_\alpha}\phi^+)
({\cal D}_\beta\phi)]\>.\eqno(37)$$

To get the luminosity $L(n)$ with respect to the Hawking radiation for TIC
with the Chern number $n$, we define a vacuum state $|0>$ by the conditions
$$a^-_i|0>=0, \> b^-_i|0>=0\>,\eqno(38)$$
and then

$$L(n)=\lim_{r\to\infty}\,\int\limits_{S^2}\,
<0|T_{tr}|0>d\sigma \eqno(39)$$

with the vacuum expectation value $<0|T_{tr}|0>$ and the surface element
$d\sigma=r^2\sin\vartheta d\vartheta\wedge d\varphi$. Now with the help
of (34), (35), (37), (38) and passing on to the quantities $x$, $k$
introduced in Sec. 3, we shall find
$$L(n)=M^{-2}\lim_{x\to\infty}{\rm Re}\left[\frac{i}{2\pi}
\sum_{l=|n|}^\infty(2l+1)
\int_{\mu_0M}^\infty\>\overline{(\partial_x\psi(x,k,n,l)}
\psi(x,k,n,l)dk\right] \>,\eqno(40)$$
where the relations
$$\frac{d}{dr}=\frac{1}{a}\frac{d}{dr^*}=\frac{1}
{aM}\frac{d}{dx}\eqno(41) $$
were used.

As a consequence, the choice of vacuum state can be defined by the choice of
the suitable linear combination of (31) for $\psi(x,k,n,l)$. Under
this situation the Hawking radiation corresponds to the mentioned choice in
the form
$$C^1_{nl}(k)=\frac{1}{\sqrt{\exp(\frac{k^+}{TM})-1}},\>
C^2_{nl}(k)=0\>,\eqno(42)$$
which at last with using asymptotics (22)--(23) entails (in usual units)
$$L(n)=
A\sum\limits_{l=|n|}^\infty(2l+1)
\int\limits_{\tilde{\mu}}^\infty\,\frac{|s_{11}(k,n,l)|^2\,k^+dk}
{\exp(\frac{k^+}{TM})-1}\>, \eqno(43)$$
where the black hole
temperature $T$ is given by (4),
$A=\frac{1}{2\pi\hbar}\left(\frac{\hbar c^3}{GM}\right)^2\approx
0.273673\cdot10^{50}\,{\rm{erg\cdot s^{-1}}}\cdot~M^{-2}$,
$\tilde{\mu}=(\mu_0M)/\mu_{pl}^2$, $\mu_0$ and $M$ in g while
$\mu_{pl}^2=c\hbar/G$ is the Planck mass square in g$^2$.

We can interpret $L(n)$ as an additional contribution to the
Hawking radiation due to the additional charged scalar particles leaving
the black hole because of the interaction with monopoles and
the conforming radiation
can be called {\it the monopole Hawking radiation}\cite{{Gon97},{Gon99}}.
Under this situation,
for the all configurations luminosity $L$ of black hole with respect
to the Hawking radiation concerning the complex scalar field to be obtained,
one should sum up over all $n$, i. e.
$$L=\sum\limits_{n\in{\Bbb{Z}}}\,L(n)=L(0)+
2\sum\limits_{n=1}^\infty\,L(n)
\eqno(44)$$
since $L(-n)= L(n)$.

As a result, we can expect a marked increase of Hawking radiation from
black holes under consideration.
But for to get an exact value of this increase one should
apply numerical methods. In the case of the pure SW black hole
($\Lambda=0$),
for example, it was found that the contribution due to monopoles
can be of order 11~\%  of the total pion-kaon luminosity\cite{GF}. So that
it would be interesting enough to evaluate similar increase in the case
$\Lambda\ne0$.

\section{Concluding remarks}
  The results of the present paper show that nontrivial topological
properties of black holes may have essential influence on their physical
properties even when the black holes cannot be considered as
the isolated objects. Though we dwelt only upon the scalar particle case
it is beyond doubt that similar considerations will hold true for spinor
particles too if employing the results of Refs.\cite{Gon99} The correct
statement of the corresponding scattering problem for spinor particles
on the SW black hole has been only recently explored\cite{Fir01} and
one can extend the results obtained here to the spinor case.
Also the important case is that of black hole immersed in strong magnetic field.
The conforming metric is the Schwarzschild-Ernst one\cite{Er76} which is
defined on the topology \bh$\>$ as well, consequently, one should investigate
how the (internal) Dirac monopoles will behave under these conditions.
We hope to continue the work along this line in future explorations.

\section{Acknowlegments}
    The work of authors was supported in part by the Russian
Foundation for Basic Research (grant no. 01-02-17157).

\nonumsection{References}

\end{document}